\documentclass[a4paper]{jpconf}
\usepackage{graphicx}
\usepackage{slashed}
\begin{document}

\def\bb    #1{\hbox{\boldmath${#1}$}}
\def\agt   {~\raisebox{0.6ex}{$>$}\!\!\!\!\!\raisebox{-0.6ex}{$\sim$}~}
\def\alt   {~\raisebox{0.6ex}<\!\!\!\!\!\raisebox{-0.6ex}{$\sim$}~}

\title{Feynman Amplitude Approach to study the Passage
of a Jet in a Medium}

\author{Cheuk-Yin Wong}

\address{Physics Division, Oak Ridge National Laboratory, Oak Ridge, TN 37831}

\ead{wongc@ornl.gov}

\begin{abstract}
In the Feynman amplitude approach for coherent collisions of a jet
with medium partons, the Bose-Einstein symmetry with respect to the
interchange of the exchanged bosons leads to a destructive
interference of the amplitudes in most regions of the phase space but
a constructive interference in some other regions.  As a consequence,
there is a collective longitudinal momentum transfer to the scatterers along the jet
direction, each scatterer carrying a substantial fraction of the
incident jet longitudinal momentum.  The manifestation of the
Bose-Einstein interference
may have been observed in angular correlations of hadrons associated
with a high-$p_T$ trigger in high-energy collisions at RHIC and LHC.
\end{abstract}

\section{Introduction}
To study the process of multiple collisions of a jet with medium
partons, it is important to distinguish coherent and incoherent
collisions \cite{Won12}.  The scattering process between the jet and a
medium parton can be described by the exchange of a gluon.  Such an
exchange takes time.  One can infer the gluon exchange time, which can
also be called the longitudinal coherence time, from the longitudinal
momentum transfer in a single two-body collision.  In the rest-frame
of the medium, the longitudinal momentum transfer $q_z$ in each
two-body scattering is related to the transverse momentum transfer
$q_T$ and the incident jet momentum $p_{\rm jet}$ by
\begin{eqnarray}
q_z=q_T^2/2p_{\rm jet}.
\label{eq1}
\end{eqnarray}
 The corresponding longitudinal coherence time is
\begin{eqnarray}
\Delta t_{\rm coh} = \hbar / q_z c = 2 \hbar p_{\rm jet} / q_T^2 c.
\end{eqnarray}
 If $\Delta t _{\rm coh}$ is much greater than the mean-free time
 $\lambda/c$ between collisions, the exchange of one gluon is finished
 before another collision begins and the jet-(medium partons)
 collisions are incoherent.  If $\Delta t _{\rm coh} \ll \lambda/c$,
 the exchange of one gluon is not completed before another collision
 begins and these jet-(medium partons) collisions are coherent.

For a mini-jet of $p_{\rm jet}\sim 10$ GeV and a typical transverse
momentum transfer of $q_T\sim 0.4$ GeV at RHIC and LHC, we estimate
that the longitudinal coherence time $\Delta t _{\rm coh}$ is of order
25 fm/c, which is much greater than the mean-free time for multiple
collisions.  Jet-(medium parton) multiple collisions at RHIC and LHC
are coherent collisions.

Conventional investigations on coherent collisions of a fast particle
with medium scatterers use the potential model
\cite{Gla59}-\cite{Arm12} in
which the scatterers are represented by static potentials and the
longitudinal recoils of the scatterers are considered as dependent
variables that are appendages to the deflected motion of the incident
particle.   However, the coherent collisions  of $p$ with $n$ medium partons in the reaction
$p+a_1+a_2+...+a_n
\to p'+a_1'+a_2'+...+a_n'$ is  a $(1+n)$-body collision with 3($n$+1) degrees of freedom, subject to an over-all four-momentum conservation.  
 A general treatment of coherent collisions necessitates the
use of the {\it both} the transverse and longitudinal recoil momenta of the scatterers as independent
dynamical variables, which is not allowed in the potential
model.  This leads us to forgo the potential model and to turn to the
 Feynman amplitude approach for the general treatment of coherent
collisions.

The Feynman amplitude gives the probability amplitude as a function of
the initial and final particle momenta of the interacting particles.
For a set of given initial momenta, a set of medium final recoil
momenta can be reached by different paths in different Feynman
diagrams.  There are thus uncertainties in the specifications of the
vertices along $\Delta z_{\rm coh}$ at which various virtual bosons
are emitted.  By Bose-Einstein (BE) symmetry, the total Feynman amplitude is the sum of all amplitudes with all
interchanges of the virtual boson vertices.  The summation of these
amplitudes and the accompanying interference constitute the
Bose-Einstein interference in the passage of the jet in the dense
medium.

\section{Feynman Amplitude Approach and Bose-Einstein Interference}

\begin{figure} [h]
\hspace*{0.7cm}
\includegraphics[scale=1.0]{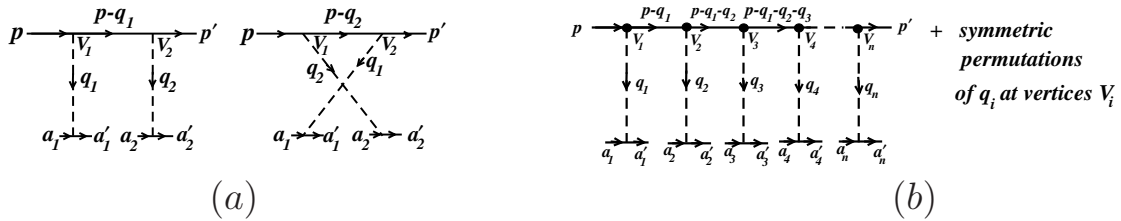}
\caption{ Feynman diagrams for the collision of a fast fermion $p$
  ($a$) with two medium fermions $a_1$, and $a_2$, and ($b$) with $n$
  medium fermions $a_1$, $a_2$,..., and $a_n$.  }
\end{figure}
As an illustration of the Bose-Einstein interference of Feynman
amplitudes, we consider first the example of the collision of a
fermion $p$ with two fermion scatterers $a_1$ and $a_2$, in $p+a_1+a_2
\to p'+a_1'+a_2'$ in the Abelian gauge theory as shown in
Fig.\ (1$a$).  For simplicity, we assume the rest masses of the
fermions to be the same and consider the high-energy limit in which $
\{ p_0, |{\bb p}| ,p_0', |{\bb p}'|\} \gg \{ |{\bb a_i }|, q_0,|{\bb
  q_i }|\} \gg m, {\rm ~for~} i=1,2$.  We assume conservation of
helicity in high-energy collisions.  The total amplitude $M$ is the
symmetrized sum of the two diagrams in Fig.\ (1$a$).  In the
high-energy limit the total amplitude $M$ is
\begin{eqnarray}
M(q_1,q_2)
\sim
g^4 \frac{2 p\cdot {\tilde a}_1 2 p\cdot {\tilde a}_2}{2m^3}
\left (
 \frac{1}{2 p \cdot  q_1 - i \epsilon} 
+\frac{1}{2 p \cdot  q_2 - i \epsilon}
\right )
\frac{1}{q_2^2}  
\frac{1}{q_1^2},
\label{eq3}
\end{eqnarray} 
\begin{eqnarray}
{\rm where} \hspace*{2.7cm}
{\tilde a}_i
=
\sqrt{\frac{a_{i0}+m}{{a_{i0}'+m} }}
\frac{a_i'}{2}
\!+\! \sqrt{\frac{a_{i0}'+m}{a_{i0}+m} }
\frac{a_i}{2}.
\hspace*{6.0cm}
\end{eqnarray}  
The fermion $p'$ is outside the interaction region and is on the mass
shell after the collision, leading to the constraint
 \begin{eqnarray}
(p-q_1-q_2)^2-m^2~\sim~ -2p\cdot q_1-2p\cdot q_2 ~\sim~ 0.
\label{6}
\end{eqnarray}
Because of this mass-shell constraint (\ref{6}), the two amplitudes in
Eq.\ (\ref{eq3}) are correlated with each other.  The real parts of
the amplitudes in Eq.\ (\ref{eq3}) destructively cancel each other,
leaving imaginary sharp distributions at $2p\cdot q_1$ and $2p\cdot
q_2$,
\begin{eqnarray}
\hspace*{-0.5cm}
M(q_1,q_2)
\!\sim\!
\frac{g^4}{2m}  \frac{2 p\cdot {\tilde a}_1}{m_1q_1^2}
\frac{ 2 p\cdot {\tilde a}_2}{m_2q_2^2 }
\biggl  \{\!
 i \pi \Delta(2 p \cdot  q_1)\!+\!i\pi \Delta(2 p \cdot  q_2)
\!\biggr \}\!,
\label{16a}
\end{eqnarray} 
\begin{eqnarray}
{\rm where}\hspace*{4.0cm}
\Delta(2p\cdot q_1)=\frac{1}{\pi}\frac{\epsilon}
{(2 p \cdot  q_1)^2+\epsilon^2},\hspace*{6.0cm}\nonumber
\end{eqnarray} 
which approaches the Dirac delta function $\delta(2p\cdot q_1)$ in the
limit $\epsilon \to 0$.

Generalizing to the case of the coherent collision of a fast fermion
with $n$ fermion scatterers in the process
$
p+a_1+...+a_n \to p' + a_1'+ ...+a_n'$, 
the total Feynman amplitude  as given in Fig.\ (1$b$) is
\begin{eqnarray}
M(q_1,q_2,...,q_n)=\frac{ g^{2n}}{2m} 
\left \{  \prod_{i=1}^n \frac{2 p\cdot {\tilde a}_i }{m_i q_i^2 } \right \}
\left \{ \prod_{j=1}^{n-1}\frac{1}
{\sum_{i=1}^j 2p\cdot q_i - i \epsilon}
+ {\rm symmetric~ permutations} \right \}.
\label{18a}
\end{eqnarray}
Using earlier results of Cheng and Wu \cite{Che87} for
 the above sum of Feynman amplitudes with symmetric
permutations, 
\begin{eqnarray}
\biggl \{ \prod_{j=1}^{n-1}\frac{1}
{\sum_{i=1}^j 2p\cdot q_i - i \epsilon}
+ {\rm symmetric~permutations}
 \biggr \} \Delta (\sum_{i=1}^n 2p\cdot q_i)
=(2\pi i)^{n-1} \prod_{i=1}^n \Delta (2p\cdot q_i),\label{18}
\end{eqnarray}
we obtain the sum Feynman amplitude $M$, including
all the symmetric permutations of the exchanged bosons  \cite{Won12}
\begin{eqnarray}
M(q_1,q_2,...,q_n)
\Delta (\sum_{i=1}^n \!2p\cdot q_i)
=\frac{ g^{2n}(2 \pi i)^{n-1}}{2m^{2n-1}}   \prod_{i=1}^n \frac{2 p\cdot {\tilde a}_i }{q_i^2 }
 \Delta (2p\cdot q_i).
\label{eq8}
\end{eqnarray}
This can be alternatively rewritten as 
\begin{eqnarray}
M(q_1,q_2,...,q_n)=\frac{ g^{2n}(2 \pi i )^{n-1}} {2m^{2n-1}n} 
 \left (  \prod_{i=1}^n 
\frac{ 2 p\cdot \tilde a_i} { q_i^2} \right )
 \sum_{j=1}^n\biggl  \{
\prod_{i=1,i\ne j}^n   \Delta (2p\cdot q_i) \biggr \}.
\label{eq7}
\end{eqnarray}
Equation (\ref{eq7}) for the Feynman amplitude as a sum of the product
of delta functions of $2p\cdot q_i$, is similar to previous 
Bose-Einstein interference results
obtained for the emission of many photons or gluons in bremsstrahlung
and for the sum of ladder and cross-ladder amplitudes in the collision
of two fermions \cite{Che69}-\cite{Lam97b}.

The above considerations for the Abelian theory can be extended to the
non-Abelian theory \cite{Won12}.  For the coherent collision of an
energetic parton with parton scatterers in non-Abelian cases, we find
that the complete Bose-Einstein symmetry in the exchange of virtual
gluons consists not only of space-time exchange symmetry but also
color index exchange symmetry.  Nevertheless, there is always a 
space-time symmetric and color-index symmetric component of the
Feynman amplitude  that behaves in the same way as the Feynman
amplitude in the Abelian case.
For this space-time symmetric and color-index symmetric component, the
recoiling partons behave in the same way as in collisions in the
Abelian case.  As the amplitude of this component involves products of
the singular delta functions, it is also the dominant component 
 and provides the dominant contribution to the 
 total Feynman amplitude.

For gluon scatterers, there is a minor modification in the
longitudinal momentum distribution with $\tilde a_i$ of quark
scatterers in Eq.\ (\ref{eq7}) replaced by \cite{Won12}
\begin{eqnarray}
\bar a_i = \frac{a_i+a_i'}{2}.
\end{eqnarray}

\section{Consequences of the Bose-Einstein Interference}

 The total Feynman amplitude $M$ in (\ref{eq7}) reveals that 
 the BE interference of the Feynman amplitudes gives rise to delta-function type constraints $\Delta (2p\cdot q_i)$, 
which impose the conditions 
\begin{eqnarray}
q_{i0}-q_{iz}=\frac{\bb p_T \cdot {\bb q}_{iT}}{p_{\rm jet}}\sim 0~~~~{\rm
  or~~~}q_{i0}\sim q_{iz}.
\label{eq9}
\end{eqnarray}
The gluon propagator $1/q_i^2$ in Eq.\  (\ref{eq7})  becomes
\begin{eqnarray}
\frac{1}{q_i^2} = \frac{1}{q_{i0}^2-q_{iz}^2
-|\bb q_{iT}|^2} \sim \frac{1}{-|\bb q_{iT}|^2}.
\label{eq10}
\end{eqnarray}
We obtain the important result that as a consequence of the BE
interference of the Feynman amplitudes, the scatterers tend to come
out at small $q_{iT}$ peaking at zero, with $q_{i0}\sim q_{iz}$.

To proceed further, we need the distribution of the longitudinal
momentum transfer $q_{iz}$ of the scatterers, which is determined by
the Feynman amplitude $M$ and the phase-space factors via the differential
cross section given by \cite{Won12}
\begin{eqnarray}
d^{n}\sigma = \frac{|M|^2(2\pi)^4  \delta^4\{
p+\sum _{i=1}^n a_i - p'-\sum_{i=1}^n a_i')}{\{\prod_{i=1}^nf_{pi}\} 
\{\prod_{i=2}^n(m/p_{i0}) \}
 \{\prod_{i=2}^n T_i\}}   
\left \{\frac{d^4{ p}'2m}{(2\pi)^3}
\frac{D_{p'}(p')}{ 2 p_{0}' }\right \}
\biggl \{ {\prod_{i=1}^n   \frac{d^4 a_i'2m_i}{(2\pi)^3}
\frac{D_i( a_{i}')}{ 2 a_{i0}}} \biggr \},~~
\label{21}
\end{eqnarray}
where  $f_{pi}$ is the flux factor for  the collision 
 between  $p_i$ and the scatterer $a_i$ of rest mass $m_i$,
\begin{eqnarray}
f_{pi}&=&\frac{4\sqrt{ (p_i\cdot a_{i})^2 - (m ~m_{iT})^2}}{(2m)(2m_i )},
\label{161}
\end{eqnarray}
and $m_{iT}=\sqrt{m_i^2+{\bb
    a}_{iT}^2}$.  
The final fast particle $p'$ resides outside the medium 
and  its state can be described as being   on the mass shell,  
\begin{eqnarray}
\frac{D_{p'}(p')}{2p_0'} \sim \frac{\delta (p_0' -\sqrt{ {\bb p' }^2 + m^2})}{2p_0'} =\delta (p'^2-m^2).
\end{eqnarray}
The states of the medium scatterer after the collision, $a_i'$,  can be described by
\begin{eqnarray}
D_i( a_{i}') =\frac{\Gamma_i/2\pi}{  [a_{i0}'-
\sqrt{(\bb a_i' - g_i\bb A)^2 + (m_i+S)^2 } +g_i A_0]^2 + \frac{\Gamma_i^2}{4} },\nonumber
\end{eqnarray}
where $A=\{\bb A, A_0\}$ and $S$ are the vector and scalar mean fields
experienced by the medium scatterer $a_i'$ after the collision,
respectively.  As the mean fields and scatterer widths increase with
medium density and are presumably quite large and dominant for a dense
medium, we shall approximately represent $D_i(a_i')$ as an average
constant that is only a weak function of $a_i'$.  Other descriptions
of $D_i(a_i')/2a_{i0}'$ for the states of the scatterers are also
possible but may not be as general; they can be the subjects for
future investigations.

In the high-energy limit, the constraint $\Delta (2p \cdot q_i)$ is
the same as the constrain $\Delta ( (p-q_i)-m^2)$ which can be
interpreted as an intermediate state $p_{i+1}$=$(p-q_i)$ appearing as
a quasi-particle on the mass shell as a result of the interference of
many amplitudes.  The quantity $T_i$ with $\{i=2,..n\}$ in Eq.\ (\ref{21})  is  the mean lifetime of the intermediate quasi-particle
state $p_i$ of the incident fast particle, prior to its exchanging a
virtual boson with $a_i$.

We change variables from $
a_i'$ to $ q_i=a_i'-a_i$, integrate over 
$ d^4 p'$, and we obtain 
\begin{eqnarray}
d^n \sigma  =\frac{1}{4}
\left (  \frac{\alpha^2}{mp_{\rm jet}} \right )^{ n}
\delta(p_0'+\sum q_{i0} -p_0)
\left \{ \prod_{i=1}^{n}\frac{8D_{i}}{f_{pi}} \frac{ (2p\cdot {\tilde a}_i)^2 }{m  2a_{i0}'} \frac{dq_{iz} d{\bb q}_{iT}}{   |\bb q_{iT}|^4} \!\!\right \}\! .
\label{22}
\end{eqnarray}
We introduce the fractional longitudinal momentum
kick
\begin{eqnarray}
x_i=\frac{q_{iz}}{p_{\rm jet}},~~~~~
dq_{iz}=p_{\rm jet} dx_i,
\end{eqnarray}
we obtain then approximately the differential cross section
\begin{eqnarray}
\hspace*{-0.5cm}d^n \sigma 
\sim 
C f(x_1,x_2,x_3,...,x_n)\delta(1-x_1 - x_2 - x_3-...-\frac{p_z'}{p_{\rm jet}})
 \frac{dx_1 dx_2 ... dx_n 
d{\bb q}_{1T}  d{\bb q}_{2T} ...d{\bb q}_{nT} }
{  |\bb q_{1T}|^4~|\bb q_{2T}|^4~...~|\bb q_{nT}|^4},
\label{48}
\end{eqnarray}
where $C$ is a weak function of $x_i$ and $q_{iT}$,
and \cite{Won12}
\begin{eqnarray}
 f(x_1,x_2,x_3,...,x_n)\sim 
\begin{cases} { ~~~~~~~ 1 & {\rm ~~~for~ quark~scatterers},\cr
                    \frac{1}{x_1 x_2 x_3 ... x_n} &
{\rm ~~~for~ gluon~scatterers}. \cr}
\end{cases}
\end{eqnarray}
The average longitudinal momentum fraction 
is approximately
\begin{eqnarray}
\langle x_i\rangle \sim 
\begin{cases}
{
\frac{1}{2n} & {\rm~ for~ quark ~scatterers}, \cr
 \frac{1-m_gT/p_{\rm jet}} 
 {n \cosh ^{-1}(p_{\rm jet}/nm_{gT})} &
{\rm~ for~ gluon ~scatterers}. \cr }
\label{48}
\end{cases}
\end{eqnarray}
Equation (\ref{48}) indicates that the ratio of the average
longitudinal momentum transfer $q_{iz}$ to the incident jet momentum
$p_{\rm jet}$ is approximately inversely proportional to $n$.  The
longitudinal momentum transfer $q_{iz}$ is much greater than $|\bb
q_{iT}|$, as the transverse momentum transfer peaks at small values of
$|\bb q_{iT}|$.  As a consequence, the angular direction of the
momentum transfer is predominantly along the jet direction.  Thus, in
a coherent collision there is a collective quantum many-body effect
arising from Bose-Einstein interference such that each scatterer
receives a substantial momentum transfer, predominantly along the jet
direction, with a magnitude that is approximately inversely
proportional to the number medium scatterers in the collision, $n$.

\section{Signatures of Bose-Einstein Interference}

The results in the above sections provide information on the signatures
for the occurrence of the Bose-Einstein interference in the coherent
collision of a jet with medium partons \cite{Won12}:

\begin{enumerate}

\item
The medium scatterers recoil collectively.

\item
Each scatterer
acquires a longitudinal momentum kick $q_z$ predominantly along the
incident jet direction
that is 
is approximately inversely proportional to
the number of scatterers $n$.

\item
The Bose-Einstein interference is 
a quantum many-body effect.  It
occurs only in the multiple collision of the fast jet with two or
more scatterers.  Therefore there is a threshold, corresponding to the
requirement of two or more scatterers in the multiple collision,
$n\ge 2$.

\end{enumerate}

\begin{figure} [h]
\hspace*{0.6cm}
\includegraphics[scale=1.0]{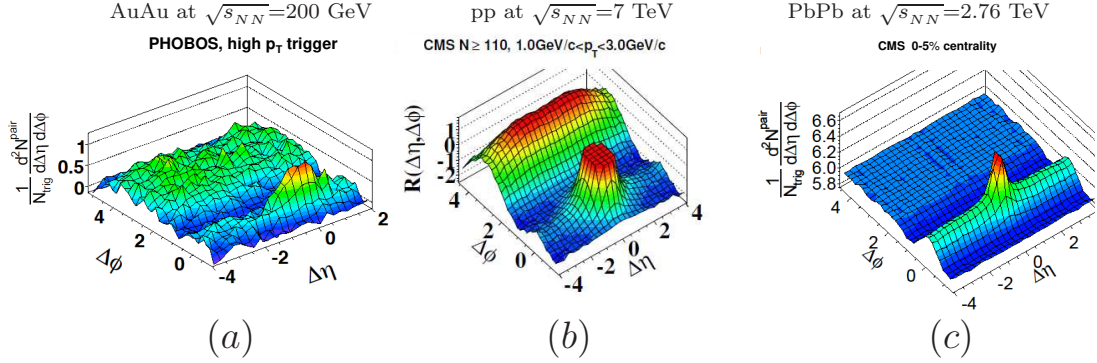}
\caption{$\Delta \phi$-$\Delta \eta$ correlation of produced hadrons
  in the most central collisions with a high $p_T$ trigger : ($a$) PHOBOS data of AuAu collisions at $\sqrt{s_{NN}}=200$ GeV at RHIC  \cite{Pho10}, ($b$)  CMS data of  $pp$ collisions at
  $\sqrt{s_{NN}}=7$ TeV at LHC  \cite{CMS10},
  and ($c$) CMS data of  PbPb collisions at $\sqrt{s_{NN}}=2.76$ TeV at LHC  \cite{CMS11}.  }
\end{figure}

To inquire whether Bose-Einstein interference may correspond to any
observable physical phenomenon, it is necessary to separate the
incident jet from the scatterers.  Such a separation is kinematically
possible in $\Delta \phi$-$\Delta\eta$ angular correlation
measurements of produced pairs with a high-$p_T$ trigger
\cite{Ada05}-\cite{ ATL12a}.  In Figs.\ 2, we show the angular
correlations with a high-$p_T$ trigger for the most central collisions : ($a$) PHOBOS data of Au-Au collisions at $\sqrt{s_{NN}}=200$ GeV at RHIC
 \cite{Pho10}, ($b$) CMS data of pp collisions
at $\sqrt{s_{NN}}=7$ TeV at LHC 
\cite{CMS10}, and ($c$) CMS data of PbPb collisions at $\sqrt{s_{NN}}=2.76$ TeV at
LHC  \cite{CMS11}.  Particles in the
``jet" part of the correlation measurement with $\Delta \eta \sim 0$
and $\Delta\phi \sim 0$ in a narrow cone can be identified as arising
predominantly from the fragmentation of a jet.  As the trigger
particle has a $p_T$ of order a few GeV and the jet collides with a
few medium particles and each scatterer  takes up about a few GeV, the
parent jet that interacts with the medium partons corresponds to a incident
particle with an initial incident momentum of $p_{\rm jet}\sim$ 10
GeV.  The ``ridge" part of the correlation measurements with $|\Delta
\eta| \agt 0.6$ and $\Delta \phi \sim 0$ can be identified as
belonging to medium partons, based on 
their rapidity plateau structure, 
their temperatures, baryon to
pion ratios, and the variations of their yields as a function of
participants \cite{Put07,Bie07,Bie07a,Lee09,Won07}.

\section{Search for the Signatures for the Occurrence 
of the Bose-Einstein
Interference}

With the ability to separate the medium scatterers as ridge particles
from the fragments of the incident high-$p_T$ jet, we can search for
the signatures for the occurrence of the Bose-Einstein interference as
listed above.  The occurrence of the Bose-Einstein interference will
be signaled by collective momentum transfers to   the scatterers (the ridge
particles)  predominantly  along the jet  direction.  Such a collective recoil will
lead to the $\Delta \phi \sim 0$ correlation of the ridge particles
with the high-$p_T$ trigger.  Indeed, as one can observe in angular
correlations of produced hadrons at RHIC and LHC shown in Fig. 2, the
ridge particles are correlated with the incident trigger jet particles
at $\Delta \phi \sim 0$, supporting the first signature of the
Bose-Einstein interference.

To investigate the second signature of the BE interference, we need to
have a measure of the number of scatterers $n$ and the magnitude of
the longitudinal transfer $q_z$. We can encode $\langle n\rangle$ and
$\langle q_{z} \rangle$ in a phenomenological momentum kick model to
analyze the experimental correlation data
\cite{Won07}-\cite{Won11}.  The
momentum kick model identifies ridge particles as medium partons and
assumes that these medium partons have an initial rapidity plateau
distribution as a result of the string fragmentation process, and when
they suffer a collision with a jet produced in the collision, they
receive a momentum kick $q_z$ along the jet direction.  The model has
been successful in describing extensive sets of triggered associated
particle data of the STAR, PHENIX, PHOBOS, and CMS Collaborations,
over large regions of $p_t$, $\Delta \phi$, and $\Delta \eta$, in many
different phase space cuts and $p_T$ combinations, including
dependencies on centralities, dependencies on nucleus sizes, and
dependencies on collision energies
\cite{Won07}-\cite{Won11}.
  
The magnitude of the longitudinal momentum kick $q_z$ affects the width of the
ridge as a function of $\Delta \phi$ whereas the number of the kicked
medium partons $n$ gives the over-all height of the associated ridge yield.
Using the momentum kick model, experimental data for 
AuAu collisions at $\sqrt{s_{NN}}=200$ GeV at RHIC reveal that
$q_z$$\sim$0.8$-$1.0 GeV  and
$f_R \langle n\rangle$$\sim$3.8
 for the most central collisions, 
where
$f_R$ is the attenuation factor for ridge particles.  If we take $f_R$
to be the same as the attenuation factor $f_J$ for jet fragment
particles, with $ f_J\sim 0.63$ obtained by comparing jet fragment
yields in AuAu and $pp$ collisions, then $q_z$$\sim$0.8$-$1.0 GeV and
$ \langle n \rangle$$\sim$6 for the most central AuAu collisions
at $\sqrt{s_{NN}}=200$ GeV.  In another momentum
kick model analysis for the highest multiplicity $pp$ collisions at
$\sqrt{s_{NN}}=7$ TeV at the LHC, the momentum kick model gives
$q_z$$\sim$2 GeV and $f_R \langle n \rangle \sim 1.5$ corresponding
approximately to $\langle n \rangle$$ \sim$2.4.
The experimental data give an average longitudinal momentum transfer
$q_z$ that is approximately inverse proportional to the number of
scatterers $\langle n \rangle$, in approximate agreement with the
second signature of the Bose-Einstein interference.

\begin{figure} [h]
\hspace*{2.0cm}
\includegraphics[scale=0.75]{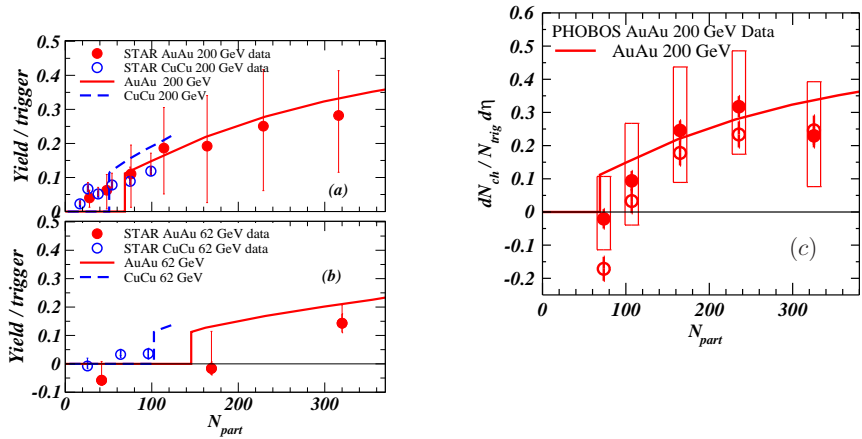}
\caption{ The ridge yield per high-$p_T$ trigger as a function of the
  participant number $N_{\rm part}$ for nucleus-nucleus collisions at
  $\sqrt{s_{NN}}$=200 and 62 GeV \cite{Sta12,Pho10}.  
  The curves are the momentum kick model results of \cite{Won08a}
  modified to include the Bose-Einstein interference threshold effect
  of $n \ge 2$. }
\end{figure}

The third signature for the occurrence of the Bose-Einstein
interference is the presence of a threshold at $n$=2.  The collective
recoil of the scatterers along the jet direction occurs only when the
number of scatterers exceed two.  This signature will indicate a
sudden onset of the ridge yield as a function of an increase in the number
of kicked medium scatterers, $n$.  The number of kicked medium
scatterers $n$ increases with centrality, as can be represented
either by an increase in the number of participants $N_{\rm part}$, or by
an increase in the multiplicity number $N_{\rm mult}$.  We need a
description of the number of medium parton scatterers $n$, the participant number $N_{\rm part}$, or  the 
multiplicity number $N_{\rm mult}$,  as a function of centrality.

Previously, we have worked out the number of medium parton scatterers,  the number of participants, and parton
energy loss,  as a function of centrality for nucleus-nucleus collisions using a geometrical model of
jet-(medium parton) collisions \cite{Won08a}.  The basic assumption is a
jet-medium parton cross section, the systematics of the density of the
hot medium, and the behavior of the fragmentation function as a
function of energy loss.  The available data of $R_{AA}$ as a function
of centrality and the number of medium scatterers at the most central
collision as inferred from the correlation data provide the needed reference data to determine the centrality dependency of the number of
medium scatterers.  In the momentum kick model, as it is assumed that
the associated particle yield is proportional to the number of
scatterers, the momentum kick model can predict the centrality
dependency of the associated ridge yield.

We show in Fig.\ 3 and the experimental ridge yield per high-$p_T$
trigger as a function of $N_{\rm part}$ for AuAu and CuCu collisions
at $\sqrt{s}=200$ and $62$ GeV at RHIC \cite{Bie07,Nat08,Pho10}.  We
also show in Fig.\ 3 the theoretical yields obtained in the momentum
kick model \cite{Won08a}, where the ridge yield at the most central
collision at $N_{\rm part}$=320  for AuAu collision at $\sqrt{s_{NN}}=200$ GeV
was calibrated as $n=6$ \cite{Won08a}.  With
such a calibration, the threshold values in $N_{\rm part}$ at which
$n=2$ can be located.   Theoretical ridge yields from the momentum
kick model analysis, modified to include the Bose-Einstein interference
threshold effect of $n \ge 2$, are shown as the solid curves for AuAu
collisions, and as dashed curves for CuCu collisions in Fig.\ 3.  The
theoretical thresholds in Fig.\ 3 will be smoothed out by the
uncertainties in the estimates of the number of scatterers and the
fluctuations of the number of scatterers as a function of $N_{\rm
  part}$.  Although the experimental data appear to be consistent with
theory and with the presence of thresholds, the large error bars in
the STAR data in Figs.\ 3($a$) and 3($b$) preclude a definitive
conclusion.  On the other hand, the PHOBOS measurement as shown in
Fig. 3($c$) indicates more clearly a threshold at $N_{\rm part}\sim
70$, in agreement with the predicted $n \agt 2$ threshold for the
occurrence of the collective medium parton recoil that is a signature
for the occurrence of the BE interference.

\begin{figure} [h]
\hspace*{1.8cm}
\includegraphics[scale=0.58]{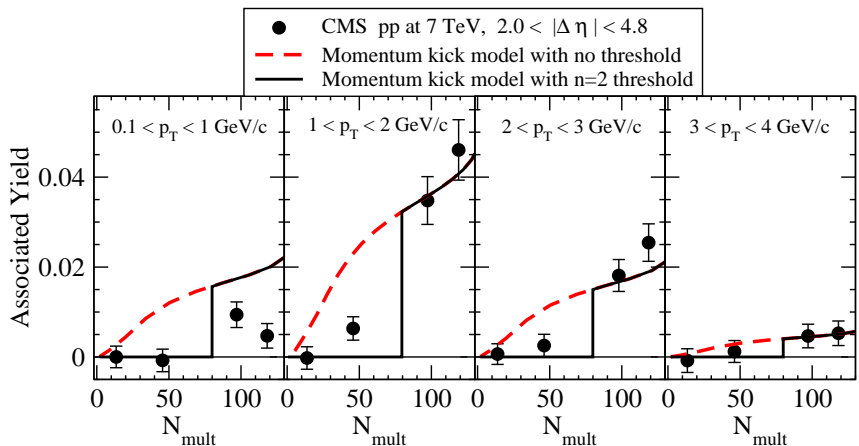}
\caption{CMS Collaboration data \cite{CMS10} for $pp$ collisions at
  $\sqrt{s_{NN}}=7$ TeV for different $p_T$ regions of associated
  particles.  The solid and dashed curves are the results of the
  momentum kick model, with or without a threshold at $n=2$.  }
\end{figure}

For the ridge yield as a function of multiplicity in $pp$ collisions,
it is necessary to determine the centrality dependence of both the
ridge yield and the charge multiplicity.  In the momentum kick model,
the ridge yield is proportional to the number of kicked medium
partons.  The (average) number of kicked medium partons per jet,
$f_R\langle n ({\bb b}) \rangle$, can be evaluated as a function of
the impact parameter ${\bb b}$ for $AA$ collisions \cite{Won08a,Won09}
.  We apply a similar analysis to $pp$ collisions by treating the
colliding protons as extended droplets as in the Chou-Yang model
\cite{Cho68}.  For the multiplicity number of produced particles as a
function of centrality, we evaluate the participant droplet elements as a
function of centrality and assume that the multiplicity is
proportional to the participant droplet elements.  By combining both number
of kicked medium partons and the multiplicity number as a function of
centrality, we obtain the momentum kick model results of the
associated ridge yield as a function of multiplicity shown in Fig.\ 4.
The data points are from the CMS Collaboration for $pp$ collisions at
$\sqrt{s_{NN}}$= 7 TeV.   The dashed curves are the momentum kick
model results without a threshold, and  the solid
curves are the results of the momentum kick model with the assumption
of a threshold at $n=2$.  Results in Fig. 4 indicate the possible
presence of a ridge threshold at $n=2$, in suggestive support of the
third signature for the occurrence of the Bose-Einstein interference.

\section{Summary and Conclusions}

A general treatment of coherent collisions necessitates the use of {\it both} the transverse and the longitudinal 
recoil momenta of the scatterers as independent dynamical variables.
In the potential model \cite{Gla59,Gyu94}, however, the longitudinal
momenta transfers cannot be independent dynamical variables.  This
leads us to forgo the potential model and to use the Feynman amplitude
approach for coherent collisions.

In the Feynman amplitude approach for the coherent collisions of a
fast particle on $n$ scatterers, there are $n!$ different orderings in
the sequence of collisions along the fast particle path at which
various virtual bosons are exchanged.  By Bose-Einstein symmetry, the
total Feynman amplitude is the sum of the $n!$ amplitudes for all
possible interchanges of the virtual bosons.  The summation of these
$n!$ Feynman amplitudes and the accompanying interference constitute
the Bose-Einstein interference in the passage of the fast particle in
the dense medium.

Our interest in examining this problem has been stimulated by the
phenomenological successes of the momentum kick model in the analysis
of the angular correlations of hadrons produced in high-energy
heavy-ion collisions
\cite{Won07}-\cite{Won11}. We seek a
theoretical foundation for the origin of the longitudinal momentum
kick along the jet direction postulated in the model.  We find that in
the coherent collisions of an energetic fermion with $n$ fermion
scatterers at high energies in the Abelian theory, the symmetrization
of the Feynman scattering amplitudes with respect to the interchange
of the exchanged bosons leads to the Bose-Einstein interference,
resulting in a sharp distributions at $p\cdot q_i\sim 0$.  The
Bose-Einstein symmetry constraints of $p\cdot q_i\sim 0$ limit the
transverse momentum transfers of the scatterers to small values of
$q_{iT}$.  The longitudinal momenta of the scatterers get their share
of longitudinal momenta from the jet, resulting in the collective momentum transfer to the scatterers along the jet direction.

For coherent collisions of an energetic parton with parton
scatterers in non-Abelian cases, we find that the complete
Bose-Einstein symmetry in the exchange of virtual gluons consists not
only of space-time exchange symmetry but also color index exchange
symmetry.  Nevertheless, there is always a space-time symmetric and
color-index symmetric component of the Feynman amplitude that behaves
in the same way as the Feynman amplitude in the Abelian case.
Furthermore, as these amplitudes involve products of singular delta
functions, they are also the dominant components.  There is thus a
finite probability for the parton scatterers to acquire a momentum transfer 
along the trigger jet direction, each carrying a significant
fraction of the longitudinal momentum of the incident jet.
 The collective recoil has a threshold, 
requiring at least two scatterers
 for the occurrence of the quantum interference.  Theoretical analysis of
the magnitude of the Feynman diagram matrix elements and phase space
factors indicates that the average magnitude of the momentum kick should be 
approximately inversely proportional to the number of scatterers.

In high-energy nuclear collisions at RHIC and LHC, an incident
high-$p_T$ jet and the scatterers can be kinematically separated by
angular correlation measurements.  A comparison of the jet
and ridge components reveals the presence of a  collective momentum transfer to  the scatterers along the jet direction, and a
phenomenological analysis using
the momentum kick model supports  the approximate
inverse proportionality relation between the magnitude of the longitudinal momentum kick and the number of scatterers.  The empirical 
ridge yield as a function of centrality suggests the possible presence of  a threshold at $n\sim 2$, as
required for the quantum Bose-Einstein interference.  Thus, the
manifestation of the Bose-Einstein interference effects may have been
experimentally observed in the $\Delta \phi$-$\Delta\eta$ angular correlation of
hadrons associated with a high-$p_T$ trigger in high-energy 
collisions at RHIC and LHC.

\vspace*{0.7cm}
\noindent{\bf Acknowledgment}

The author would like to thank Profs. Vince Cianciolo, Horace
W. Crater, C. S. Lam, and Jin-Hee Yoon for helpful discussions. This
research was supported in part by the Division of Nuclear Physics, U.S. Department of Energy.


\end{document}